\documentstyle[preprint,aps]{revtex} 
\begin{document}
\draft
\title{\bf {Revisiting Thouless conductance formula }}
\author{P. Singha Deo\cite{eml} }
\address{Institute of Physics, Bhubaneswar 751005, India}
\maketitle
\begin{abstract}
It was shown using perturbation theory[1] that Thouless energy
Ec for a quantum system scales linearly with the conductance of
the system. We derive in an alternate way in 1-D that Ec scales
with the conductance in a very different way. We physically show
the difference between our approach and that of ref. 1 to expect
our results to hold in higher dimensions also. We verify our
results with exact numerical calculations.
\end{abstract}
\pacs{PACS numbers :72.10.F,61.72,71.10}
\narrowtext
\newpage

Quantum mechanical (electron) transport through disordered system
has been of great importance for quiet a long time. A major
simplification of the problem was conceived by Landauer when he
proposed that although resistance is an outcome of dissipation
and breakdown of time reversal symmetry one can formulate
resistance in terms of elastic scattering only if one assumes a
clear spatial separation between elastic processes and inelastic
processes[2].  The so called Landauer's conductance formula is
now on firm grounds and is at the heart of studying transport
through mesoscopic systems experimentally as well as
theoretically[3].  Another important idea was put forward by
Thouless[1]. According to him extended states are very sensitive
to twisting of boundary conditions and also contribute highly to
conductance. Whereas localized states are not so sensitive to
twisting of boundary conditions and also contribute very little
to conductance. So there must be a relation between conductance
and a characteristic energy scale of the system called Thouless
energy Ec. In his original work[1] Ec is defined as the average
change in the eigen-energies of the system if we twist the
boundary conditions from periodic to anti-periodic.
Ec so defined appears to be a very
artificial energy scale but it was latter found to be the
typical inverse diffusion time[4]. This energy scale appears as a
bridge to our understanding of universality from Anderson
localization to Quantum chaos, universal conductance
fluctuations[5], persistent currents[6,7] and many more. Using
perturbation theory it was shown that Ec scales linearly with
conductance[1] in the diffusive regime. In the localized regime
in all dimensions (including 1D) Ec was found to be zero.
However as Thouless relation is only for a finite sample (for an
infinite sample average level spacing $\Delta$ goes to zero)
there is no need to rule out the possibility of a relation
between Ec and the conductance G even in the localized regime,
i.e., when the localization length is much smaller than the
sample length. The fact that Ec turns out to be zero when using
perturbation theory is due to a particular approximation as
stated in ref. 1. In this treatment we find a way to avoid this
approximation in 1D and thus we find a scaling between Ec and G
in the localized regime. In fact the relation derived here is
general to all regimes (ballistic, diffusive and localized).
Again diffusive regime in 1D means the localization length is
much larger than the sample length.

Recent experiments on persistent currents has made it necessary
to re-examine the Thouless formula critically[7].  In ref. 1
unless we appeal to Kubo Greenwood formula in 3D the results are
true in all dimensions. One can just as well appeal to
Landauer's conductance formula in 1D. The fact that the scaling
between Ec and G may not be linear in 1D when transmission is
very small was first pointed out by Anderson and Lee[8]. But
this work starts from the eigenvalues of the S matrix and M
matrix and not from the eigen-functions and eigenvalues of the
Hamiltonian as in ref.1 and hence it is not obvious whether it
is a specialty of 1D as stated in ref. [8].  Recent work[9]
however shows that Ec is linearly related to G as claimed in ref
1. We start from the eigen-functions and so the difference with
ref. 1 is evident.  Hence we can argue physically to expect our
results to hold in higher dimensions also. We find that the
scaling in 1D to the first order is same as that of Anderson and
Lee. We write down all higher order terms in terms of the
actually measurable two probe conductance and so in our case
transmission need not be small. Our result is general to all
three regimes.

As we have already mentioned that unless one appeals to Kubo
formula in 3D the treatment of ref. 1 is identical in all
dimensions we start by repeating its steps in 1D. Let there be a
system of size L with a random potential V(x) defined by the
Hamiltonian H and the eigen-functions $\phi(x)$ satisfy periodic
boundary conditions. This means $\phi(x)$ is the eigen-function
of an infinite periodic potential formed by the repetition of
the sample of length L. Change in boundary condition is
equivalent to a gauge transformation such that the wave function

\begin{equation}
\psi(x)=\phi (x)e^{i\alpha x}
\end{equation}

\noindent where $\phi(x)$ satisfy periodic boundary condition
and satisfies the Schrodinger equation with additional terms
$\alpha^2 +2 \alpha p_x$, where $p_x$ is the momentum operator.
As $\phi(x)$ satisfy periodic boundary condition anti-periodic
boundary condition for $\psi(x)$ is obtained by putting $\alpha
L=\pi$ and periodic boundary condition for $\psi(x)$ is obtained
by putting $\alpha L=0$.

So $\alpha^2+2\alpha p_x$ can be treated as a perturbation to
the Hamiltonian with periodic boundary conditions whose
eigen-functions are $\phi(x)$ and thus Ec can be calculated
according to its original definition. However the shift is
identified with the lowest order non-zero term in the
perturbative expansion. We call this lowest order term $Ec_l$
and

\begin{equation}
Ec_l=\alpha^2 + 2 \alpha \Sigma_j {\mid [p_x]_{ij} \mid ^2 \over
E_i-E_j}
\end{equation}

Here $[p_x]_{ij}$ are the matrix elements of $p_x $ between
$\phi(x)$. Ref. 1 assumes these terms to be exponentially small
and this assumption makes $Ec_l$ go to zero in the localized
regime.  However $\phi(x)$ is not an eigen-function of a random
potential but the eigen-function of an infinite periodic
potential formed by the repetition of the sample of length L.
Hence $\phi(x)$ is not an exponentially decaying localized state
but an extended Bloch state however random V(x) is. Hence the
neglected terms are not exponentially small but are reasonably
large. We shall soon discuss that the difference can be as large
as 100 times the actual value. After this one can  relate $Ec_l$
to the conductance through Kubo Greenwood formula (in the
diffusive regime) through the matrix elements of $p_x$. However
in Kubo formula the matrix elements of $p_x$ is calculated
between wave-functions of the system with open boundary
conditions (coupled to a bath) to account for dissipation[9,10].
Hence these states are not extended Bloch states for a random
V(x) as that in eqn (2). Besides the approximation (used in ref.
[1]) of the diffusive spectra being uncorrelated is not
appropriate[9,11].

We start from eigen-functions given in eqn(1) and this
allows us to express $Ec$ in terms of conductance.

\begin{equation}
\psi(x+L)=\phi (x+L) e^{i\alpha (L+x)}
\end{equation}

Now $\phi(x)$ is the eigen-function of a periodic potential and
satisfies periodic boundary conditions. So we can write
$\phi(x+L)=\phi(x)$ and thus we explicitly take care of the
fact that the states are extended Bloch states. Hence

\begin{equation}
\psi(x+L)=\phi (x) e^{i\alpha x}
e^{i\alpha L}=\psi(x) e^{i\alpha L}
\end{equation}

Now as a consequence of Bloch's theorem[12] we can get eqn 19 of
ref 7.

\begin{equation}
\alpha L=\cos^{-1}re[1/t(E)]
\end{equation}

Where t(E) is the transmission amplitude at an incident energy
E, E being the eigen-energy corresponding to the Bloch
eigen-function $\phi(x)$. RHS of eqn 5 is the phase of the
wave-function of the Bloch state. This could not be realized in
ref. 7. When the conductor becomes perfectly ordered eqn 5
becomes $\alpha$L=kL, k being the momentum. In presence of
disorder eqn 5 is just $\alpha$L=KL K being the Bloch momentum,
the state E being an extended state.  This is exact for any
arbitrary V(x) and we do not have to consider ballistic,
diffusive or localized regime separately.

We can write $re[{1\over t(E)}]={\cos(\beta)\over \mid t \mid}$
where $\beta$ is the transmission phase.  Then we can take
$\cos(\beta)=\cos{2\pi E \over \Delta}$. For a clean system
where the dispersion is E$\equiv k^2$ this is exact.  Except in
the limit of extremely strong disorder this assumption is fairly
accurate[13,14]. Now the bound-states of the system can be easily
found using the periodic and anti-periodic boundary conditions.

The bound states $E_n^p$ for the periodic boundary condition is
obtained by putting $\alpha$=0, i.e.,

\begin{equation}
\cos^{-1}[{cos({2\pi E \over \Delta})\over \mid t(E)\mid}]=0
\end{equation}

\noindent and the bound-states $E_n^{ap}$ for the anti-periodic
boundary condition is obtained by putting $\alpha =\pi$/L, i.e.,

\begin{equation}
\cos^{-1}[{cos({2\pi E \over \Delta})\over \mid t(E)\mid}]=\pi
\end{equation}

\noindent and solving eqns 6 and 7 for E.  Hence the bound-states
for the periodic boundary condition are

\begin{equation}
{2\pi E_n^p \over \Delta}=\cos^{-1} \mid t(E_n^p) \mid
\end{equation}

\noindent The bound-states for the anti-periodic boundary
condition are

\begin{equation}
{2\pi E_n^{ap} \over \Delta}=\cos^{-1} (-\mid t(E_n^{ap})) \mid
\end{equation}

Then we use the expansion[15]

\begin{equation}
{\pi \over 2}-\cos^{-1}(y)=y+{1\over 2.3}y^3+{1.3\over 2.4.5}
y^5+ {1.3.5\over 2.4.6.7} y^7+....\,\,\,for\,\, y^2\le 1
\end{equation}

to obtain

$$2\pi {\mid E_n^p-E_n^{ap} \mid\over \Delta}=((\mid t(E_n^p)
\mid+\mid t(E_n^{ap})\mid)+{1\over 2.3}(\mid t(E_n^p) \mid^3+ \mid
t(E_n^{ap}) \mid^3)+$$
\begin{equation}
{1.3\over 2.4.5}(\mid t(E_n^p) \mid^5 +\mid
t(E_n^{ap})\mid^5+.....)
\end{equation}

Now Landauer's conductance formula gives the dimensionless
conductance g(E) as

\begin{equation}
g(E)=\mid t(E) \mid ^2
\end{equation}

Now we can take the average of both sides of(11) over disorder
configuration. In 1D where we do not have a mobility edge it is
appropriate to take an arithmetic mean of the LHS. It is also
appropriate to take the arithmetic mean of the RHS because $\mid
t(E) \mid$ is less than unity. Thus we get Ec/$\Delta$ in terms
of dimensionless conductance g(E) as given by Landauer's
formula. Note that $<\mid t(E_n^p) \mid>=<\mid t(E_n^{ap})
\mid>$ because modulus of transmission do not depend on the
phase of the wave function.

\begin{equation}
{Ec\over \Delta}=1/\pi(< \surd{g(E)} >+{1\over 2.3}<
\surd{g(E)} >^3+
{1.3\over 2.4.5}< \surd{g(E)} >^5+.....)
\end{equation}

We have nowhere invoked the condition that transmission is
small.  It shows to a leading order Thouless energy depends on
$<\surd{g(E)} >$ and not $< g(E)>$ as obtained in ref. [8].
Thouless energy $Ec$, i.e., the average energy difference
between a periodic and anti-periodic system is a fundamental
energy scale governing transport in a quantum system. $Ec$ that
appears in eqn. 13 is not just the shift given by the first
order perturbation but is the shift if all higher order
terms are taken into account. We can relate this fundamental
energy scale to the conductance. This may shed light on the
nature of transport in a random medium more accurately.

However the shift as given by the lowest order perturbation term
i.e., $Ec_l$ can be very accurately calculated from the
curvature of energy versus $\alpha L$ dispersion curve at
$\alpha L$=0. Subsequently we find the scaling between $Ec_l$
and $g$. From eqn 5

\begin{equation}
\cos{2\pi E \over \Delta}=\mid t(E) \mid \cos(\alpha L)
\end{equation}

For small $\alpha L$

\begin{equation}
{2\pi E \over \Delta}=\cos^{-1}[\mid t(E) \mid w]
\end{equation}

where $w=1-(\alpha L)^2/2+(\alpha L)^4/4+....$. Again using the
expansion given in eqn 10 we find

$${2\pi E \over \Delta}=\pi /2 - \mid t(E) \mid -{1 \over 2.3}
\mid t(E) \mid^3 - {1.3 \over 2.4.5} \mid t(E) \mid^5 -....$$
$$+\mid t(E) \mid {(\alpha L)^2 \over 2} + {1 \over 2.3} \mid
t(E) \mid^3 {3 (\alpha L)^2 \over 2} + {1.3 \over 2.4.5} \mid
t(E) \mid^5 {5 (\alpha L)^2 \over 2} + .....$$
\begin{equation}
a^\prime (\alpha L)^4 + b^\prime (\alpha L)^6+....
\end{equation}

$a^\prime$,  $b^\prime$, etc. are functions of $\mid t(E) \mid$
whose explicit forms we do not need to know.

Then using (12) and taking average of both sides over different
disorder configurations one finds

\begin{equation}
{2 \pi \over \Delta} {d^2 E \over d(\alpha L)^2} \mid_{(\alpha
L)=0}={2 \pi \over \Delta} Ec_l=\surd g(E) + (\surd g(E))^3/2
+(\surd g(E))^5/2.4 + ....
\end{equation}

Again we find that to a leading order $Ec_l$ goes as $\surd g$
and not as g. As we can write down all higher order terms our
expression is valid for entire range of $\mid t \mid$. This is
the extra benefit of our alternate derivation. Ref [1]
underestimates the shift because the eigenstates used in the
perturbative calculations were not taken as Bloch states. As
soon as we imply periodic boundary condition we impose a
discrete symmetry in the system that results in making the
eigenstates of the system extended Bloch states however strong
the disorder is.

Recent experiments on persistent currents support this. Magnetic
field can be taken as a physical realization of $\alpha L$ and
persistent currents (averaged over disorder and summed over
levels) arising due to sensitivity of eigenstates to twisting of
boundary condition can be taken as a measure of Ec.  Ref. [7] is
the only theory known so far that gives the magnitude of
persistent current correct to an order of magnitude.  Ref [7]
shows that in a ring with radial grain boundaries states are
Bloch states. In such a situation disorder can suppress
conductance 100 times more than it suppresses persistent
currents. Whereas in a ring with point defects states are
localized and then disorder suppresses persistent currents by
the same amount it suppresses conductance. When one calculates
the ratio between two numbers the proportionality constants do
not matter at all but the scaling matters. This suggests
localized states will show linear scaling between Ec and G and
the treatment of Thouless goes through.

In higher dimensions if we start with periodic boundary
conditions in all three directions (as is the original
definition of $Ec$) then we imply the same discrete symmetry as
in 1-D and the states are extended. Twisting boundary condition
in one direction will shift the states more than that given by
perturbation theory.  However in higher dimension we do not know
a simple expression for Bloch states in terms of conductance as
in (5).  One can numerically study these situations although one
may have to resort to more complicated averaging procedures.

This result is expected to hold for the tight binding model also
although the coefficients can change a lot. In the continuum
model average $\Delta$ as well as average conductance
monotonically increase with energy but not in the tight binding
model.  These will change the coefficients drastically but the
same type of scaling of Ec with g should be there. This we
verify numerically.  Also in the limit of extremely strong
disorder $\Delta$ approximately goes as W/N[14] where W is the
strength of the disorder and N is the number of sites. Hence the
above equation can be extended to the regime of extreme strong
disorder too.

To verify  numerically for a tight binding chain we take a chain
consisting of 15 sites. We find its eigenvalues for periodic as
well as anti-periodic boundary conditions by exact numerical
diagonalization. We arrange the eigen values in ascending order
and then calculate average Ec/$\Delta$.  Then we average over
100 disorder configurations.  $<\surd g>$ is also evaluated
exactly using transfer matrix method, and then averaging over
100 configurations. In fig (1) we show that Ec/$\Delta$ depends
on $<\surd g>$ in a way that can be best fitted to a polynomial
of the type given in eqn (13). A minimum of three terms are
needed to give a reasonable fit suggesting the importance of
higher order terms.  In fig.  (2) we show that Ec/$\Delta$
versus $<\surd g>$ cannot be fitted to a power law curve
although the power in the best fit is close to one is very
suggestive. Also a fit to a general polynomial of three terms is
much worse than the fit in fig.  (1) if one keeps in mind that
after all it is a three parameter fit. Other type of fits also
do not work.  We have left out the extreme strong disorder limit
where the fit deteriorates. However one can study this regime
with the modification explained before.

Hence Thouless argument leading to eqn. (13) and (17) will give
the conductance as that given by Landauer's conductance formula
in all three regimes of 1D. $Ec_l$ and $\Delta$ can be
calculated using diagonalization techniques and in some cases
eqn. (17) may be easier to deal with. It may specially simplify
the problem of treating the effect of e-e interaction on
conductance.

The author thanks Professors A. M. Jayannavar and S. M.
Bhattacharjee for discussions.

\vfill
\eject

\vfill
\eject
FIGURE CAPTIONS

Fig. 1. Plot of Ec/$\Delta$ versus $<\surd g>$ shown by dots.
Solid line is the best fit to a polynomial of the type given in
eqn. (13) which is y=1.84691 x - 2.9037 $x^3$ + 2.04372 $x^5$.

Fig. 2. Plot of Ec/$\Delta$ versus $<\surd g>$ shown by dots.
Solid curve 1 is the best fit to a power law curve which is
y=.902978 $x^{1.1292}$. The solid curve 2 is the best fit to a
general polynomial of three terms which is y=4.81251 x -
9.62012 $x^2$ + 5.78701 $x^3$.

\vfil
\eject


\begin{thebibliography}{99}
\bibitem[*]{eml} prosen@iopb.ernet.in
\bibitem{1} J. T. Edwards and D. J. Thouless, J. Phys. C. {\bf
5}, 807(1972).
\bibitem{2} R. Landauer, Z. Phys. {\bf 68} 217(1987).
\bibitem{3} A. D. Stone and A. Szafer, IBM J. Res. Dev. {\bf 32}
384(1988); H. U. Baranger and A. D. Stone, Phys. Rev. B {\bf 40}
8169(1989); J. Noeckel, H. U. Baranger and A. D. Stone, Phys.
Rev. B {\bf 48} 17569(1993).
\bibitem{1a} D. J. Thouless, Phys. Rep. {\bf 13}, 93(1974).
\bibitem{3} B. L. Altshuler and B. D. Simons, Universalities:
from Anderson localiztion to Quantum Chaos, Elsevier Sc.
Publishers(1994).
\bibitem{5} M. B$\ddot u$ttiker, Y. Imry and R. Landauer, Phys.
Lett. A 96 (1983) 365.
\bibitem{4} G. Kirczenow to appear in Journal of Physics,
Condensed matter; P. Singha Deo submitted to Journal of Physics,
Condensed matter.
\bibitem{and} P. W. Anderson and P. Lee, Suppl. of the Prog. of
Thoe. Phys. {\bf 69} 212(1980).
\bibitem{ak} E. Akkermans and G. Montambaux, Phys. Rev. Lett.
{\bf 68}, 642(1992).
\bibitem{nt} N. Trivedi and D. A. Browne, Phys. Rev. B, {\bf
38}, 9581(1988); R. Landauer and M. B$\ddot u$ttiker, Phys. Rev.
Lett. {\bf 54}, 2049(1985).
\bibitem{shl} B. I. Alt'shuler and B. Shklovskii, Zh. Eksp.
Teor. Fiz. {\bf 91}, 220(1986).
\bibitem{5}D. W. L. Sprung, Hua Wu, J. Martorell, Am. J.
Phys. {\bf 61}, 1118(1993). P. Singha Deo and A. M.
Jayannavar, Phys. Rev. B. {\bf 50} 11629(1994); Solid State
Physics by N. W. Ashcroft and N. D. Mermin, eqns. 8.76 and 8.80.
\bibitem{6} M. L. Meheta, Random Matrices (Academic, New York,
1967); U. Shivan and Y. Imry, Phys. Rev. B {\bf 35},
6074(1987).
\bibitem{6} H. Cheung, Y. Gefen, E. K. Riedel, W. Shih,
Phys. Rev. B 37 (1988) 6050.
\bibitem{10} Table of Integrals, Series, and Products, I. S.
Gradstein and I. M. Ryzhik, 5th edition, pg 59.
\end{thebibliography}
\end{document}